\documentclass[aps,prl,showpacs,notitlepage,nofootinbib,superscriptaddress,floatfix,showkeys,twocolumn]{revtex4-1}
\usepackage{dcolumn}
\usepackage{blindtext}
\usepackage{hyperref}
\usepackage{amsmath,amssymb}
\usepackage{float}
\usepackage{microtype}
\usepackage{graphicx}
\usepackage{bm}
\usepackage{latexsym}
\usepackage{epsfig}
\usepackage{psfrag}
\usepackage{color}
\usepackage[dvipsnames]{xcolor}
\usepackage{subfigure}
\usepackage[section]{placeins}
\usepackage{natbib}
\usepackage{ulem}
\usepackage{lipsum}
\usepackage{comment}

\def\nn{\nonumber} 
\def\f{\frac}

\def\l{\left}
\def\r{\right}
\def\d{{\mathrm{d}}}

\def\HI{H_{\mathrm{I}}}

\def\ee{\eta_{\mathrm{e}}}
\def\e1i{\epsilon_{1\mathrm{i}}}

\def\HI{H_{\mbox{\tiny{I}}}}

\def\As{A_{\mbox{\tiny{S}}}}

\def\nb{n_{\mbox{\tiny{B}}}}

\def\pb{\mathcal{P}_{\mbox{\tiny{B}}}}

\def\pe{\mathcal{P}_{\mbox{\tiny{E}}}}
\def\pbi{\mathcal{P}_{\mbox{\tiny{B}}}^{\mbox{\tiny{I}}}}
\def\pei{\mathcal{P}_{\mbox{\tiny{E}}}^{\mbox{\tiny{I}}}}

\def\ksb{k_{\mbox{\tiny{SB1}}}}
\def\kssb{k_{\mbox{\tiny{SB2}}}}
\def\rhogw{\rho_{\mbox{\tiny{GW}}}}
\def\ogw{\Omega_{\mbox{\tiny{GW}}}}

\def\ogwp{\Omega_{\mbox{\tiny{GW}}}^{\mbox{\tiny{PRI}}}}
\def\ogws{\Omega_{\mbox{\tiny{GW}}}^{\mbox{\tiny{SEC}}}}
\def\Ppri{\mathcal{P}_{\mbox{\tiny{PRI}}}}
\def\Psec{\mathcal{P}_{\mbox{\tiny{SEC}}}}
\def\Psecinf{\mathcal{P}^{\mathrm{inf}}_{\mbox{\tiny{SEC}}}}
\def\Ppriinf{\mathcal{P}^{\lambda,\mathrm{inf}}_{\mbox{\tiny{PRI}}}}
\def\Psecre{\mathcal{P}^{\lambda,\mathrm{re}}_{\mbox{\tiny{SEC}}}}
\def\Psecra{\mathcal{P}^{\lambda,\mathrm{ra}}_{\mbox{\tiny{SEC}}}}

\def\fnb{f_{n_{\mbox{\tiny{B}}}}}

\def\Fnb{\mathcal{F}_{n_{\mbox{\tiny{B}}}}}

\def\pce{\mathcal{P}_{\mbox{\tiny{E}}}^\mathrm{c}}
\def\pcb{\mathcal{P}_{\mbox{\tiny{B}}}^\mathrm{c}}

\def\nw{n_w}
\def\ke{k_{\mathrm{e}}}
\def\Mpl{M_{\mbox{\tiny{Pl}}}}
\def\MP{M_{\mbox{\tiny{Pl}}}}
\def\wre{w_{\mathrm{re}}}
\def\Tre{T_{\mathrm{re}}}

\def\Nre{N_{\mathrm{re}}}

\def\kre{k_{\mathrm{re}}}

\def\knu{k_{\nu}}
\def\gre{g_{\mathrm{re}}}
\def\rhoe{\rho_{\mathrm{e}}}

\def\ae{a_{\mathrm{e}}}
\def\are{a_{\mathrm{re}}}

\def\mB{\mathcal{B}}

\def\xe{x_{\mathrm{e}}}
\def\l{\left}
\def\r{\right}
\def\xre{x_\mathrm{re}}

\def\gre{g_{\mathrm{re}}}
\def\cH{\mathcal{H}}

\def\rhoe{\rho_{\mathrm{e}}}

\def\vk{\bm k}
\def\vq{\bm q}

\def\ee{\eta_{\mathrm{e}}}
\def\enu{\eta_{\nu}}
\def\ere{\eta_{\mathrm{re}}}
\def\xnu{x_{\nu}}
\def\knu{k_{\nu}}

\def\Sk{\mathcal{S}_{\bm k}}

\def\Gk{G_k}
\def\Gkinf{G_k^{\mathrm{inf}}}
\def\Gkre{G_k^{\mathrm{re}}}
\def\Cmre{\mathcal{C}_{\mathrm{m}}^{\mathrm{re}}}

\def\rhoemi{\rho_{\mathrm{em}}^{\mathrm{I}}}

\def\dneff{\Delta N_{\mathrm{eff}}}

\def\kmin{k_{\mathrm{min}}}

\def\Gkra{G_k^{\mathrm{ra}}}
\def\mCra{\mathcal{C}_{\rm ra}}
\def\mCre{\mathcal{C}_{\rm re}}
\def\Cmra{\mathcal{C}_{\rm r}^{\rm ra}}

\def\Ptcs{\mathcal{P}_{\rm T,s}^{\rm c}}

\def\GeV{\mathrm{GeV}}

\allowdisplaybreaks[1]

\begin{document}

\title{Constraining inflationary magnetogenesis and reheating 
via GWs in light of PTA data}
\author{Subhasis Maiti}
\email{E-mail: subhashish@iitg.ac.in}
\affiliation{Department of Physics, Indian Institute of Technology, Guwahati, 
Assam, India}
\author{Debaprasad Maity}
\email{E-mail: debu@iitg.ac.in}
\affiliation{Department of Physics, Indian Institute of Technology, Guwahati, 
Assam, India}
\author{L.~Sriramkumar}
\email{E-mail: sriram@physics.iitm.ac.in}
\affiliation{Center for Strings, Gravitation and Cosmology, Department of 
Physics, Indian Institute of Technology Madras, Chennai~600036, India}
\begin{abstract}
By leveraging the limits on primordial magnetic fields (PMFs), their 
contributions to secondary gravitational waves (GWs), and the recent 
observations by the pulsar timing arrays (PTAs), we arrive at constraints 
on the epoch of reheating. 
We find that the combined spectral energy density of primary and secondary
(generated by the PMFs) GWs can be described as a broken power law with 
different indices.
We show that PMFs with blue spectra and appropriate reheating scenarios 
can successfully explain the PTA observations without invoking any new 
physics.
\end{abstract}
\maketitle


\noindent
\underline{\it Introduction:}\/~Observing cosmic magnetic fields and gravitational 
waves (GWs) has been a long-standing endeavor with resounding successes. 
Magnetic fields of the order of~$\mu\mathrm{G}$, with coherence lengths 
of tens to hundreds of Kpc, have been observed in galaxies and clusters of 
galaxies~\cite{Turner:1987bw,Grasso:2000wj,Giovannini:2003yn,Kronberg:2001st}. 
The recent $\gamma$-ray observations~\cite{MAGIC:2022piy} indicate that the 
intergalactic voids could host weak magnetic fields of strength $10^{-16}\,
\mathrm{G}$ with a coherence length as large as~Mpc~\cite{Takahashi:2011ac,
Arlen:2012iy}.
In addition, the anisotropies in the cosmic microwave background (CMB) provide an 
upper bound on the strength of the primordial magnetic fields~(PMFs) to be of the 
order of $\mathrm{nG}$ on Mpc scales~\cite{paoletti2022constraints,Zucca:2016iur}. 
On the other cosmological front, the detection of GWs is a relatively recent
phenomenon. 
Over the last few years, the LIGO-Virgo observatories have detected GWs from the 
coalescences of a large number of compact binaries~\cite{LIGOScientific:2016aoc}.  
Moreover, the latest, 15-year data from different Pulsar Timing 
Arrays (PTAs)---viz. NANOGrav~\cite{NANOGrav:2023gor}, European PTA (EPTA)
[including data from the Indian PTA (InPTA)]~\cite{2023arXiv230616224A}, 
Parkes PTA (PPTA)~\cite{Reardon:2023gzh}, and the Chinese 
PTA~(CPTA)~\cite{Xu:2023wog}---suggest the presence of a stochastic background
of GWs in the nano-Hertz (nHz) range of frequencies.

Inflation provides an efficient mechanism for the generation of GWs~\cite{Starobinsky:1979ty,
Grishchuk:1974ny,Guzzetti:2016mkm} as well as PMFs~\cite{Ratra:1991bn,PhysRevD.37.2743,
Ferreira:2013sqa,Subramanian:2015lua,Campanelli:2008kh,Jain:2012jy,Caprini:2014mja,
Sharma:2018kgs, Bamba:2021wyx} from the quantum vacuum. 
However, the details of the inflationary origin of magnetic fields and the effects 
due to post-inflationary dynamics remain uncertain. 
The detection of GWs with distinct spectral signatures induced by the PMFs could
provide insight into both the dynamics during inflation and magnetogenesis. 
Although prior studies have explored these interactions, they have often been 
carried out in simplified scenarios~\cite{Sharma:2019jtb, Caprini:2014mja,
Ito:2016fqp,1971ApJ...163..255P,Sorbo:2011rz}.

In this letter, we investigate, for the first time, the spectral energy density~(SED) 
of GWs induced by the PMFs across a generic history of reheating. 
Our results show that the phase of reheating produces unique features in the SED of 
GWs with multiple breaks across various frequency ranges, that are, in principle, 
detectable by the future GW observatories. 
These distinct features can be clearly differentiated from the SED of GWs generated 
by other sources. 
We show that the SED of GWs generated in certain reheating scenarios can 
explain the recent observations of a stochastic GW background (SGWB) detected 
by the PTAs in the nHz range of frequencies (in this context, see for instance, 
Refs.~\cite{2023arXiv230617124L,2023arXiv230617149F,2023arXiv230617822C,
2023arXiv230700646D,Liu:2023pau,liu2023implications,HosseiniMansoori:2023mqh,
zhao2023exploring, Zhu:2023gmx, Vagnozzi:2023lwo, Vagnozzi:2020gtf, Liu:2023pau, Liu:2023ymk, HosseiniMansoori:2023mqh}).

Despite a variety of cosmological observations favoring 
inflation~\cite{Planck:2015fie,Planck:2018jri,Planck:2018vyg}, direct 
evidence is still lacking.
With the advent of newly proposed, state-of-the-art observatories that aim 
to observe the SGWB~\cite{LIGOScientific:2016aoc,
LIGOScientific:2016jlg,Sathyaprakash:2012jk,Baker:2019pnp,Suemasa:2017ppd,
Amaro-Seoane:2012aqc,Barausse:2020rsu,Janssen:2014dka,NANOGrav:2023gor,
2023arXiv230616224A,Reardon:2023gzh,Xu:2023wog}, our findings could play an 
an important role in simultaneously constraining inflation, 
reheating~\cite{Haque:2021dha,Vagnozzi:2020gtf,Vagnozzi:2023lwo,Benetti:2021uea}, 
and primordial magnetogenesis~\cite{PhysRevD.82.023511,PhysRevD.98.103525, 
PhysRevD.107.043531,PhysRevD.37.2743,Ferreira:2013sqa,Subramanian:2015lua,
Campanelli:2008kh,Jain:2012jy,Caprini:2014mja,Sharma:2018kgs,Bamba:2021wyx}.   


\noindent
\underline{\it Background dynamics during reheating:}\/~In the standard cosmological 
model, inflation is followed by the phase of reheating, during which energy is 
transferred from the inflaton to radiation. 
This phase is characterized by the mass of the inflaton, its self-coupling and 
coupling to radiation. 
In the perturbative regime, these parameters can be mapped to the equation-of-state
(EoS) parameter~$\wre$ and the reheating temperature~$\Tre$. 
During reheating, the energy density of the inflaton evolves in terms of the scale
factor $a$ as $a^{-3(1+\wre)}$.
The reheating temperature, defined when the energy density of the inflaton equals
the energy density of radiation, is given by~\cite{PhysRevLett.113.041302}
\begin{align}
\Tre = \left(\frac{90 \HI^2 \Mpl^2}{\pi^2 \gre}\right)^{1/4} 
\exp \left[ -\frac{3(1+\wre)\Nre}{4} \right],
\end{align}
where~$\Nre$ is the duration of reheating counted in terms of $e$-folds, $\HI$
is the Hubble parameter during inflation, and $\Mpl \simeq 2.4 \times 10^{18}\,
\mathrm{GeV}$ is the reduced Planck mass.
Also, $\gre = 106.75$ is the effective number of relativistic degrees of freedom 
that contribute to the energy density of radiation at the end of reheating.
Note that $\HI = \sqrt{r \As}\pi\Mpl$, where $\As$ and $r$ denote the amplitude 
of the primordial scalar perturbations and the tensor-to-scalar ratio.
We shall assume that $\As = 2.1 \times 10^{-9}$ and $r_{0.05} \leq
0.036$~\cite{Planck:2018jri,Planck:2018vyg}. 
These constraints imply an upper bound on the Hubble scale during inflation, 
viz. $\HI \leq 10^{-5} \Mpl$, which we shall use in our analysis.


\noindent
\underline{\it Post-inflationary evolution of PMFs:}\/~At a conformal 
time~$\eta$, the stochastic PMFs, say, $B_i$, can be characterized by the
SED, say, $\pb(k,\eta)$. 
This SED is defined in terms of the two-point correlation function in 
Fourier space through the relation~\cite{Durrer:1999bk} 
\begin{align}\label{eq:def_em_pow}
\langle B_i^{}({\bm k},\eta) B^{*}_j({\bm k}',\eta)\rangle
= \delta^{(3)}({\bm k}-{\bm k}') P_{ij}(\hat{\bm k}) 
\f{2\pi^2}{k^3}\,\pb(k,\eta),
\end{align}
where $(i,j)$ represent the spatial indices, $\hat{\bm k}$ is the unit wave
vector along the direction of propagation, and $P_{ij}(\hat{\bm k})=\delta_{ij}
-\hat{k}_i \hat{k}_j$ is the transverse projection tensor.  
In this work, we shall assume that, at the end of inflation (corresponding to 
the conformal time~$\ee$ and the scale factor~$\ae$), the SED of
the magnetic field is given by $\pbi(k)=\mB^2(k/\ke)^{\nb}$,
where, evidently, the quantities $\mB$ and $\nb$ represent the magnitude and 
spectral index of the magnetic field and $\ke$ denotes the wave number that 
leaves the Hubble radius at the end of inflation at~$\ee$~\cite{Subramanian:2015lua,
Kobayashi:2014sga,Haque:2020bip,Tripathy:2021sfb}.
We shall further assume that the magnetic field is the dominant component 
when compared to the electric fields so that, post-inflation, the magnetic 
field evolves as $\pb(k,\eta)=\pbi(k)(\ae/a)^4$~\cite{Kobayashi:2014sga}.
We shall work with parameters $\mB$ and $\nb$ so that the PMFs do not lead to 
backreaction on the background.

Given that the fluctuations in the temperature of the CMB are of the order of 
$\delta T/T \simeq 10^{-5}$, upon assuming that they are induced by the PMFs,
it is feasible to arrive at a rough upper bound on the present-day strength 
of the magnetic field, say, $B_0^{\mbox{\tiny{{CMB}}}}$.
We find that the upper bound can be expressed as
\begin{align}
B_0^{\mbox{\tiny{{CMB}}}}
\simeq 11.7\l(\f{\HI}{10^{-5}\Mpl}\r)f_{\nb}^{1/2}
\l(\f{k}{\ke}\r)^{\nb/2}\,\mathrm{nG},\label{eq:b0-cmb}
\end{align}
where $\fnb=\nb/[1-(\kmin/\ke)^{\nb}]$ and $k_\mathrm{min}$ corresponds
to the largest observable scale today.
For a scale-invariant spectrum of PMFs (i.e. when $\nb=0$), the above value 
turns out to be $B_0^{_{\mathrm{CMB}}} \simeq 1.55\, \mathrm{nG}$.
Interestingly, this estimate proves to be close to the typical upper bound at the 
scale of $1\,\mathrm{Mpc}$ arrived at from a detailed analysis of the imprints
of the PMFs on the CMB~\cite{paoletti2022constraints, Zucca:2016iur,Planck:2015zrl,
BICEP2:2017lpa}.
In our discussion below, we shall neglect the nonlinear evolution of the 
magnetic field due to MHD processes over scales within the Hubble 
radius during the epoch of radiation domination~\cite{PhysRevLett.128.221301,
2023arXiv230617124L,2023arXiv231207938Y}.


\noindent
\underline{\it Primary and secondary GWs:}\/~In our analysis, we shall take into
account the primary GWs generated during inflation and the secondary GWs induced 
by the PMFs during inflation as well as post-inflation.
We shall consider situations wherein the generated GWs are consistent with the 
constraints from the CMB as well as big bang nucleosynthesis (BBN).
We shall compare the strengths of the induced GWs with the sensitivities of
the different GW observatories and, importantly, understand the implications
for the early universe in light of the latest PTA observations.

After magnetogenesis during inflation, we shall assume that the conformal 
invariance of the action describing the electromagnetic field is restored. 
In such a case, post-inflation, there are two stages wherein the anisotropic 
stress associated with the PMFs contributes to the generation of secondary GWs. 
The first stage corresponds to the epoch of reheating. 
The second stage is during the epoch of radiation domination until the decoupling
of neutrinos which occurs around $T_{\nu} \simeq 1\, \mathrm{MeV}$.
After decoupling, the anisotropic stress of the free-streaming neutrinos
cancel the anisotropic stress of the PMFs~\cite{Lewis:2004ef}.

Recall that the tensor perturbations, say, $h_{ij}(\eta,{\bm x})$, evolving in 
a Friedmann universe can be decomposed in terms of the Fourier modes, say, 
${h}_{\bm k}^{\lambda}(\eta)$, as follows:
\begin{equation}
h_{ij}(\eta,{\bm x})
=\sum_{\lambda=(+,\times)}\int \f{\d^3{\bm k}}{(2\pi)^{3/2}} 
e_{ij}^{\lambda}({\bm k}) h_{\bm k}^{\lambda}(\eta) 
\mathrm{e}^{i\bm{k}\cdot{\bm x}},
\end{equation}
where $e^{\lambda}_{ij}({\bm k})$ is the polarization tensor corresponding
to the mode with wave vector~${\bm k}$ and the index~$\lambda$ represents 
the two types of polarization of the GWs.
Note that $e^{\lambda}_{ij}({\bm k})$ is real in the linear polarization
basis that we shall work with.
Hence, the fact that $h_{ij}(\eta,{\bm x})$ is real implies that $h_{-\bm k}^{\lambda
}(\eta)= {h}_{\bm k}^{\lambda *}(\eta)$, and the mode functions $h_{\bm k}^\lambda(\eta)$ 
satisfy the following inhomogeneous equation~\cite{Sorbo:2011rz,
Caprini:2014mja,Sharma:2019jtb}:
\begin{equation}\label{gweq}
h_{\bm k}^{\lambda{}''}+2\f{a'}{a}h_{\bm k}^{\lambda}{}'
+k^2h_{\bm k}^{\lambda}=\mathcal{S}_{\bm k}^\lambda,
\end{equation}
where the overprimes denote differentiation with respect to $\eta$, and the
source term $\mathcal{S}_{\bm k}^\lambda$ is given by
\begin{align}
\mathcal{S}_{\bm k}^{\lambda}(\eta)
&=-\f{2}{\Mpl^2}e^{ij}_{\lambda}({\bm k})
\int \f{\d^3\vq}{(2\pi)^{3/2}} [E_i(\vq,\eta)E_j(\vk-\vq,\eta)\nn\\
&\quad +B_i(\vq,\eta)B_j(\vk-\vq,\eta)].    
\end{align}
The homogeneous solutions to this differential equation represent the primary
tensor perturbations that arise from the quantum vacuum during inflation. 
The inhomogeneous solutions correspond to the secondary tensor perturbations 
induced by the PMFs.
The power spectrum of the tensor perturbations for a given polarization 
$\lambda$ is usually defined as
\begin{align}\label{eq15}
\sum_{\lambda=(+,\times)}\l\langle h_{\bm k}^{\lambda}(\eta) 
h_{\bm k'}^{\lambda \ast}(\eta)\r\rangle
=\f{2\pi^2}{k^3}\mathcal{P}(k,\eta)\delta^{(3)}({\bm k}+{\bm k}').
\end{align}
Due to their distinct origins, the total spectrum of GWs can be expressed
as the sum of the primary and secondary components as $\mathcal{P}(k,\eta)
=\Ppri(k,\eta)+\Psec(k,\eta)$.
The spectrum of the primary GWs at any given time can be obtained by evolving
the spectrum generated during inflation which has an amplitude proportional 
to $\HI^2$~\cite{Guzzetti:2016mkm,Haque:2021dha}.
The spectrum of secondary GWs at the end of reheating (at the 
conformal time $\ere$) can be expressed as follows (see, for instance,
Ref.~\cite{Okano:2020uyr}):
\begin{align}
\Psec(k,\ere)
&=\frac{2}{\Mpl^4}\l[\int_{\ee}^{\ere} \d\eta_1\f{\ae^4
G_k(\ere,\eta_1)}{a^2(\eta_1)}\r]^2\nn\\
&\times\int_0^{\infty} \f{\d q}{q}\int_{-1}^1
\f{\d\mu f(\mu,\beta)\,\pbi(q)\pbi(|\vk-\vq|)}{[1+\l(q/k\r)^2-2\mu(q/k)]^{3/2}},
\label{eq:ps-sgws}
\end{align}
where $G_k(\eta,\eta_1)$ is the Green's function associated with Eq.~\eqref{gweq}. 
Moreover, in the linear polarization basis, the quantity $f(\mu, \beta)$ is
given by $f(\mu, \beta) =2 (1 + \mu^2)(1 + \beta^2)$ with $\mu = \hat{\bm k} 
\cdot \hat{\bm q}$ and $\beta = \widehat{({\bm k}-{\bm q})} \cdot 
\hat{\bm k}~$\cite{PhysRevD.69.063006,Zucca:2016iur,Sharma:2019jtb}.


\noindent
\underline{\it SED of GWs:}\/~Let us assume that, during inflation, the 
scale factor is given by the de Sitter form as $a=-1/\HI\eta$.
In such a case, for a power law spectrum of PMFs, the spectrum of secondary
GWs induced by the PMFs, when evaluated at the end of inflation is given 
by~\cite{Caprini:2014mja,Caprini:2018mtu}
\begin{equation}
\Psecinf(k,\ee)
=2\l(\f{\HI \tilde{\mB}}{\Mpl}\r)^4 \mathcal{I}_{\rm inf}^2\l(\frac{k}{\ke}\r)^{2\nb} \Fnb(k),
\end{equation}
where $\mathcal{I}_{\rm inf}\simeq [1-(\ke/k_{*})^{\nb}]/(3\nb)$ 
and $\Tilde{\mB}=\mB \ae^2/(\ke^2 a_0^2)$ and $a_0$ is the scale factor today.
The quantity $\Fnb(k)$ is defined as
\begin{equation}\label{def-fnbe}
\mathcal{F}_{\nb}(k) 
\simeq \f{32}{3\nb}\l[1-\l(\f{k_\ast}{k}\r)^{\nb}\r]
+\f{2\alpha}{3}\l[\l(\f{\ke}{k}\r)^{2\nb-3}-1\r]
\end{equation}
with $k_\ast$ being the CMB pivot scale and~$\alpha=56/[5(2\nb-3)]$.

During reheating, the scale factor evolves as $a(\eta)=\ae\,(\eta/\ee)^{\delta/2}$, 
where $\delta={4}/{(1+3\wre)}$.
Upon using Eq.~\eqref{eq:ps-sgws}, the spectrum of secondary GWs at the end of 
reheating can be expressed as
\begin{align}
\Psec(k,\ere)=2\l(\frac{\HI\Tilde{\mB}}{\Mpl}\r)^4
\l(\frac{k}{\ke}\r)^{\alpha_1}\Fnb(k) [\Cmre(\xre,\xe)]^2,\label{Psec}
\end{align}
where we have set $\alpha_1=2(\delta+\nb-2)$ and we have introduced 
the dimensionless variable~$x=k\eta$.
The quantity $\Cmre(\xre,\xe)$ is given by
\begin{align}\label{eq:time-int}
\Cmre(\xre,\xe)=  k\,\int_{\xe}^{\xre} \d x_1
x_1^{-\delta}G^{\rm re}_{k}(\xre,x_1)
\end{align}
with $G^{\mathrm{re}}_k(\xre,x_1)$ being the Green's function for GWs during
reheating. 
The tensor power spectrum at the time of the neutrino decoupling can be arrived 
at in a similar fashion, upon utilizing the Green's function during the epoch of
radiation domination and carrying out the integral from the conformal 
time~$\ere$ to~$\eta_\nu$.

Since the wave numbers of our interest will be well within the Hubble radius
by the late stages of the radiation-dominated epoch, the subsequent evolution 
of the energy density of GWs mirrors the behavior of the energy density of 
radiation~\cite{PhysRevD.70.043011}. 
The total, dimensionless SED of primary and secondary GWs 
{\it today}\/ (i.e. at~$\eta_0$) is defined as $\ogw(k)=  \ogwp(k,\eta_0)
+\ogws(k,\eta_0)= \rhogw(k,\eta_0)/(3 H_0^2\Mpl^2)$, where $H_0$ denotes the 
present value of the Hubble parameter.
If we assume inflation of the de Sitter form, the SED of the primary GWs
{\it today}\/ can be expressed as (in this context, see, for instance, 
Ref.~\cite{Haque:2021dha})  
\begin{align}
\ogwp(k) h^2\simeq  
\frac{\Omega_{\rm r}h^2 g_{s,0}^{1/3}}{6\gre^{1/3}}\frac{\HI^2}{\Mpl^2}  
\times\left\{\begin{array}{ll} 1 & k<\kre,\\
\mathcal{D}_1\l(\frac{k}{\kre}\r)^{-\nw} & k>\kre,
\end{array}\right.\label{eq:sd-pgws}
\end{align}
where $\mathcal{D}_1\simeq \mathcal{O}(1)$ and $\nw=2(1-3\wre)/(1+3\wre)$.
Note that $g_{s,0}$ denotes the number of effective relativistic degrees of 
freedom that contribute to the entropy density of radiation, and $\Omega_{\rm r}$
represents the dimensionless energy density of radiation today.
We find that, for $\wre\leq1/3$, the dimensionless SED of secondary GWs {\it today},\/
induced by the PMFs during inflation and post-inflation, can be written as (for
details, see supplementary material)
\begin{align}
\ogws(k) h^2 
&\simeq  \frac{\Omega_{\rm r}h^2 {g^s_0}^{1/3}}{6\gre^{1/3}} 
\left(\frac{\HI \tilde{\mB}}{\Mpl}\right)^4 \left(\frac{\kre}{\ke}\right)^{2\nb}\mathcal{I}_{\rm inf}^2\,
\Fnb(k)\nn\\
&\quad \times \left\{\begin{array}{ll}
(k/\kre)^{2\nb} &\!\!  k_{*}<k<\kre,\\
(k/\kre)^{2\nb-\nw} &\!\! \kre<k<\ke.
\end{array}\right.\label{eq:sgws_wre1}
\end{align}
Whereas, when $\wre>1/3$, we find that the dimensionless SED of secondary GWs is
given by
\begin{align}
&\ogws(k) h^2 \simeq \frac{\Omega_{\rm r}h^2 g_0^{1/3}}{6\gre^{1/3}} 
\left(\frac{\HI \Tilde{\mB}}{\Mpl}\right)^4 \left(\frac{\kre}{\ke}\right)^{2(\nb+\nw)}\Fnb(k)\nn\\
& \times \left\{\begin{array}{ll}
\mathcal{A}_1 (k/\kre)^{2\nb} &\!\!  k_{*}<k<\ksb,\\
\mathcal{I}_2(k)(\kre/\knu)^2 (k/\kre)^{2\nb+2} &\!\! \ksb<k<\knu,\\
\mathcal{I}_2(k) (k/\kre)^{2\nb} &\!\! \knu < k < \kre,\\
\mathcal{A}_2 (k/\kre)^{2\nb - |\nw|} &\!\! \kre<k \ll \kssb, \\
(k/\kre)^{2\nb + |\nw|} &\!\! \kssb \ll k<\ke,
\end{array}\right.\label{eq:sgws_wre2}
\end{align}
Note that the quantities~$\mathcal{A}_1$ and $\mathcal{A}_2$ are listed in the 
supplementary material, while $\mathcal{I}_2(k) =[\gamma+\ln(k/\kre)]^2$. 
Moreover, $k_\nu\simeq 3.9\times 10^5\,\mathrm{Mpc}^{-1}$ (i.e. $f_{\nu} 
\simeq 2.2\times 10^{-10}\,\mathrm{Hz}$) denotes the wave number (frequency)
that renters the Hubble radius at the time of decoupling of the neutrinos.
Let us highlight a few points regarding the SED of primary and secondary GWs 
we have obtained the above.
When $\wre < 1/3$, total SED exhibits two spectral breaks.
The first arises when the contribution from the secondary GWs begins to 
dominate.
The second occurs at $\kre$, where the SED transitions from $k^{2\nb}$ 
to $k^{2\nb - \nw}$. 
Notably, for $\wre>1/3$, the SED of GWs exhibit as many as four distinct 
spectral breaks at $\ksb$, $k_\nu$, $\kre$ and $\kssb$.
\begin{figure}
\centering
\includegraphics[width=0.975\linewidth]{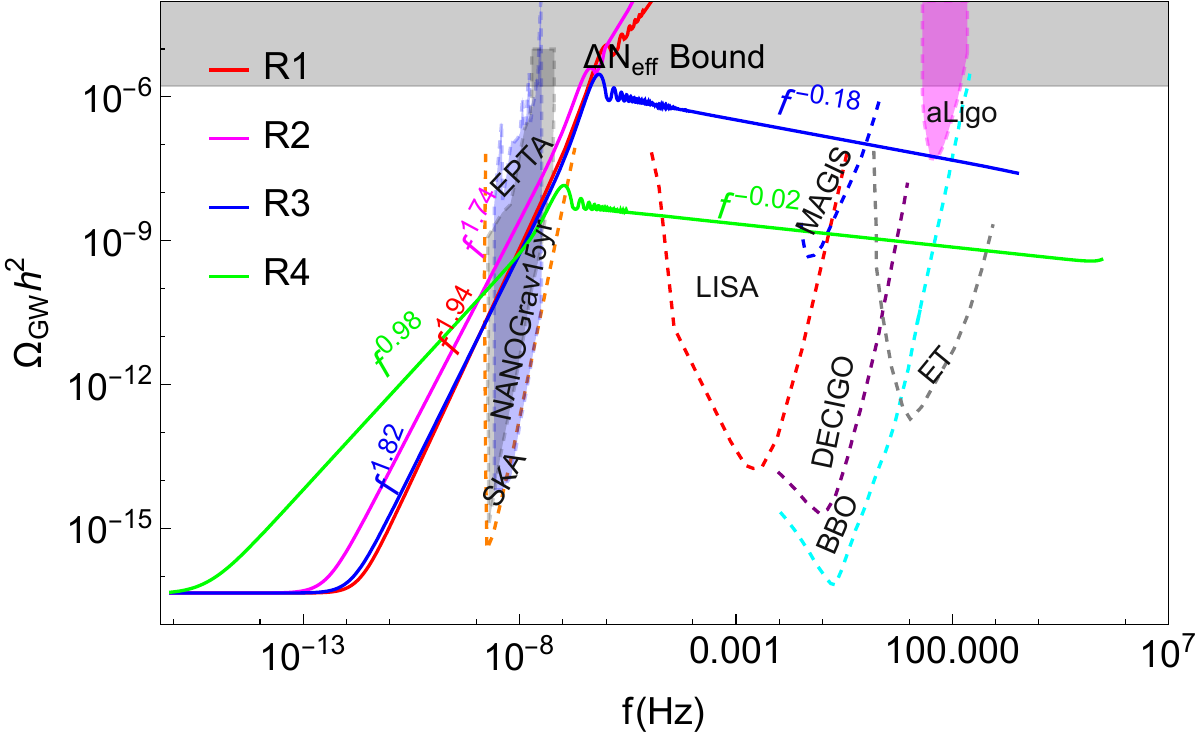}
\caption{The combined dimensionless SED of primary and secondary GWs today, i.e.
$\ogw(k) h^2$, is plotted as a function of frequency $f$ for the best-fit values
of the parameters (listed in Tab.~\ref{tab:post}) arrived at on comparing with 
the NANOGrav 15-year data.
We have plotted the SED corresponding to the best-fit values resulting from the
four runs (R1, R2, R3, R4) (in red, magenta, blue, and green, respectively).
Additionally, in the plot, we have included the NANOGrav data as well as the 
sensitivity curves of different GW wave observatories~\cite{NANOGrav:2023gor, 
Reardon:2023gzh}.}\label{fig:sd-gws}
\end{figure}

With the complete SED of primary and secondary GWs at hand,
we shall first discuss the constraints arising from the bound on~$\dneff$.
Thereafter, we shall work with parameters that are consistent with the bound
and compare the resulting SED of GWs with the PTA data.

\noindent
\underline{\it Bound on $\dneff$:}\/~When the wave numbers are inside the Hubble
radius,  as we mentioned, the energy density of primordial GWs behaves in the 
same manner as radiation.
Hence, they contribute to the effective number of relativistic degrees of freedom, 
viz. $N_\mathrm{eff}$. 
Such a behavior leads to the following constraint~\cite{Caprini:2018mtu}:
\begin{equation}
\int_{k_0}^{k_{\text{end}}} \frac{\d k}{k} \, \ogw(k) \, h^2 
\leq \frac{7}{8}\left(\frac{4}{11}\right)^{4/3}\Omega_{\gamma}h^2 \dneff,
\end{equation}
where \(\Omega_{\gamma}h^2 \simeq 2.47 \times 10^{-5}\) is the dimensionless,
present-day photon density and $\dneff$ is the observational uncertainty 
on~$N_\mathrm{eff}$~\cite{Planck:2018vyg}. 
The \textit{Planck}\/~2018 + BAO data constrain the parameter to be $N_{\rm eff} 
\simeq 2.99^{+0.34}_{-0.33}$ at $2$-$\sigma$~\cite{Planck:2018vyg}, which implies
that $\dneff \leq 0.284$.
This, in turn, leads to an upper bound of $\ogw h^2 \leq 1.6 \times 
10^{-6}$~\cite{clarke2020constraints}. 
We find that the bound restricts the parameters describing the spectrum of PMFs
as well as reheating.
For instance, if we choose $\tilde{\mB} = 1$, $\nb = 0.9$, and $\wre = 0$, we find
that the maximum reheating temperature allowed by the bound is~$\Tre\leq 0.3\,
\mathrm{GeV}$. 
Also, when $\wre > 1/3$, the maximum magnetic spectral index permitted by the 
bound turns out to be $\nb \simeq 0.35$.
However, note that the recent PTA observations indicate a SED of GWs with a strong 
blue tilt at the nHz range of frequencies.


\noindent
\underline{\it Comparison with the PTA data:}\/~
We investigate whether primordial magnetic fields (PMFs), generated during inflation via a non-conformal coupling and with conformal invariance restored at its end, could produce the stochastic gravitational wave background (SGWB) observed by pulsar timing arrays (PTAs). We compute the spectral energy density (SED) of secondary gravitational waves (GWs) sourced by PMFs during and after inflation, up to neutrino decoupling, and include primary GW contributions. Assuming low reheating temperatures, PTA-sensitive frequencies correspond to modes reentering the Hubble radius during reheating. In this case, fitting PTA data requires a magnetic spectral index \(n_{\mathrm{B}}\) close to unity and implies present-day magnetic field strengths far below blazar-inferred lower limits. We constrain the PMF and reheating parameters \((\mB, \nb, \wre, \Tre)\) by comparing the total GW SED with NANOGrav 15-year data using \texttt{PTARCade}~\cite{mitridate2023ptarcade}. While multiple parameter scans were performed (see supplementary material), we present results from four representative runs.

In our first run (R1), we kept all the parameters free and we have listed the 
resulting best-fit values of the parameters as well as the Bayesian factor in
Tab.~\ref{tab:post}.
\begin{table*}[t!]
\begin{tabular}{|c|c|c|c|c|c|c|c|c|c|c|c|}
\hline
Run 
& \multicolumn{2}{c|}{$\log_{10}(\tilde{\mB})$} 
& \multicolumn{2}{c|}{$\nb$} 
& \multicolumn{2}{c|}{$\wre$}& \multicolumn{2}{c|}{$\log_{10}(\Tre/\text{GeV})$} 
& $B_0\,\mathrm{G}$ & Bayesian & $\dneff$\\
\cline{2-9}
& Prior & Mean & Prior & Mean & Prior & Mean & Prior & Mean & & factor&  bound\\
\hline
\cline{1-11}
R1 
& $[0,5]$ & $3.1^{-0.37}_{-0.41}$ 
& $[0,1.5]$ & $0.96^{+0.18}_{-0.19}$& $[0,1.0]$ & $0.16^{+0.11}_{-0.10}$ 
& $[-2,2]$ & $1.17^{+0.24}_{-0.47}$ 
& $4.12\times10^{-20}$ & $33.38\pm7.8$ & $\times$\\
\hline
R2 
& $[0,5]$ & $3.10^{+0.73}_{-0.82}$  
& $[0,1.5]$ & $0.90^{+0.21}_{-0.20}$ & $[0.333]$ & $0.333$ 
& $[-2,1]$ & $-0.55^{+0.27}_{-0.09}$ 
& $3.34\times 10^{-16}$& $15.36\pm 4.43$ & $\times$\\
\hline
R3 
& $[0,5]$ & $3.03^{+0.77}_{-0.62}$ 
& $[0,1.5]$ & $0.91^{+0.21}_{-0.20}$ & $[0]$ & $0$ & $[-2,1]$ 
& $-0.51^{+0.27}_{0.09}$ 
& $2.53\times 10^{-25}$ & $18.65\pm 7.34$ & \checkmark\\
\hline
R4 
& $[0,5]$ & $3.32^{+0.71}_{-0.43}$ 
& $[0,0.5]$ & $0.49^{+0.05}_{-0.08}$ & $[0.1]$ & $0.1$ & $[-2,1]$ 
& $-1.15^{+0.95}_{-0.59}$ 
& $8.7\times 10^{-18}$ & $2.85\pm 0.47$ &\checkmark\\
\hline
\end{tabular}
\caption{The priors on the four parameters, viz. $(\mathcal{B},\nb,\wre,\Tre)$, 
the best-fit values arrived at on comparison with the NANOGrav data as well as 
the Bayesian factor when compared to the SMBHB model, has been listed for four 
of the runs that we have carried out.
We have also indicated if the total SED of primary and secondary GWs corresponding  
to the best-fit values are consistent with the bound on~$\dneff$.}\label{tab:post}
\end{table*}
Note that the Bayesian factor compares the cosmological source of GWs with the 
popular astrophysical scenario involving mergers of supermassive black hole binaries
(SMBHBs) for generating GWs in the nHz range of frequencies.
We find that consistency with the PTA data requires $\wre \simeq 0.16$, $\Tre
\simeq 10\,\mathrm{GeV}$ and that PMFs have a strong blue tilt with $\nb > 0.5$.
In Fig.~\ref{fig:sd-gws}, we have plotted the total SED of primary and secondary 
GWs for the best-fit values.
While the above set of values for the parameters leads to a significant Bayesian 
factor as far as the PTA data is concerned, it is clear from the figure that, for 
$f > f_{\rm re}$, the blue tilt of the SED leads to a violation of the bound 
on~$\dneff$.

\begin{figure}[t]
\includegraphics[width=0.95\linewidth]{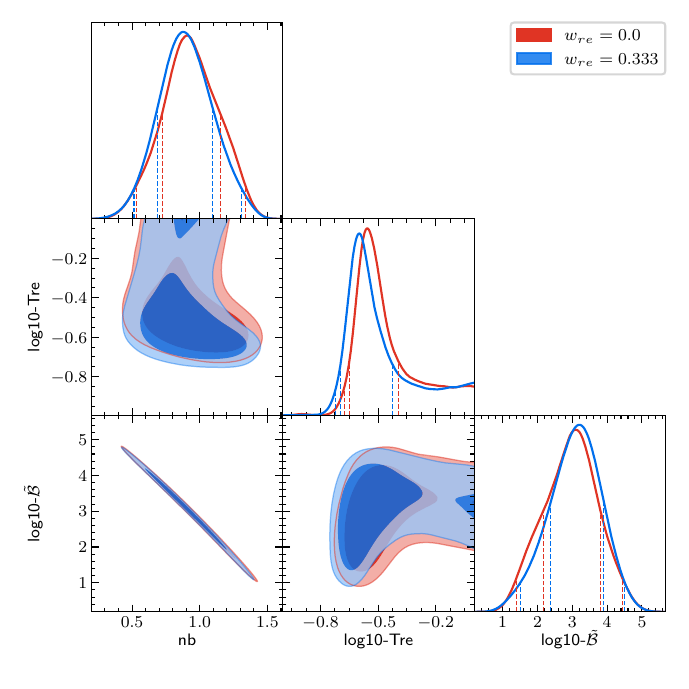}
\caption{The posterior distribution of the three parameters in the cases of 
runs R2 and R3.}\label{fig:sd-pos}
\end{figure}

In the second run (R2), we fixed $\wre=1/3$ and varied the other three parameters.
In Fig.~\ref{fig:sd-pos}, we have illustrated the posterior distributions on the 
three parameters. 
We obtain the mean value of the spectral index of the PMFs, the reheating 
temperature and the current strength of the magnetic field to be $\nb = 0.89$, 
$\Tre=0.28\,\mathrm{GeV}$ and $B_0 \simeq 3.34 \times 10^{-16}\, \mathrm{G}$,
respectively.
We find that the Bayesian factor, while it is significant, it is not as much as 
in the case of the first run (R1).
Moreover, as in the case of the first run, the total SED of primary and secondary 
GWs cross the $\dneff$ bound.

In the third run (R3), we set $\wre=0$ while varying the other three parameters. In Fig.~\ref{fig:sd-pos}, we have illustrated the posterior distributions for this case as well.
We find that the best-fit value of the spectral index of the PMFs turns out to be
$\nb = 0.91$.
We should mention that, in the range of frequencies that NANOGrav is sensitive to,
this leads to a spectral index for the SED of secondary GWs to be $n_{_{\mathrm{GW}}} 
= 1.82$.
This corresponds to an index of $\gamma_{_{\mathrm{CP}}}=5-2\,n_{_{\mathrm{GW}}} = 3.2$
for the timing residual cross-power spectral density, which is consistent with the 
estimates by the NANOGrav team~\cite{NANOGrav:2023gor}. 
Lastly, the best-fit value of $\mathcal{B}$ corresponds to a present-day magnetic 
field of $B_0 \simeq 2.53 \times 10^{-25}\, \mathrm{G}$.
Also, the Bayesian factor in this case proves to be comparable to the second run.
Importantly, since the SED of primary GWs at high frequencies (corresponding to $k>
\kre$) have a strong red tilt when $\wre=0$, we find that the total SED of primary 
and secondary GWs in this case is consistent with the $\dneff$ bound.
However, due to the strong dilution during the matter-like reheating phase and a 
low reheating temperature, we find that the strength of the present-day magnetic 
field is rather small.
Such a small value for the strength of the magnetic field today seems inconsistent
with lower bounds of $B_0 \simeq 10^{-16}\,\mathrm{G}$ as is suggested by the 
Fermi/LAT and HESS observations of TeV blazars (in this context, see 
Refs.~\cite{Neronov:2010gir,Tavecchio:2010mk,Dolag:2010ni,Dermer:2010mm,Vovk:2011aa,
Taylor:2011bn,Takahashi:2011ac}).
This issue could be resolved in scenarios where the magnetic field is generated 
after inflation, which we leave for future studies.

In the fourth and final run (R4), we set $\wre=0.1$ and vary the other three 
parameters, viz. $\mathcal{B}$, $\nb$ and $\Tre$.
In the case of $\nb$, we restrict ourselves to the range of $0<\nb<0.5$.
It should be clear from Tab.~\ref{tab:post} that, while the best fit values of 
$\mathcal{B}$ and $\Tre$ suggest a larger value of the magnetic field today (viz. 
$B_0\simeq 10^{-18}\,\mathrm{G}$) and the SED is consistent with the $\dneff$ 
bound, the fit to the PTA is not as good as with the other runs.

\noindent
\underline{\it Conclusions:}\/~Our understanding of the epoch of reheating remains
largely obscured due to the dearth of direct observational probes. 
Similarly, there exist limited constraints on PMFs. 
The recent observations by the PTAs suggesting a stochastic GW background provide
a window to constrain primordial physics on small scales.
In this work, taking into account the contributions due to the primary GWs as well 
as the secondary GWs induced by the PMFs, we have compared the total SED of GWs
with the NANOGrav 15-year data.
If we assume that the reheating temperature is greater than, say, $\Tre \simeq 1\,
\mathrm{GeV}$, then the scales corresponding to the frequencies associated with 
the NANOGrav observations fall in the domain $k_\nu < k <\kre$.
Over such a range of wavenumbers or frequencies, the SED of secondary GWs behave 
as $k^{2\nb}$ and, we find that, in order to fit the NANOGrav 15-year data, we 
require a spectral index of the magnetic field~$\nb$ that is close to unity.
Since the spectral index is high, the data favors a small value for the strength
of the magnetic field.
Moreover, given these conditions, if the SED of primary and secondary GWs is to
be consistent with the bound on $\dneff$, then we require a relatively smaller 
value of the EoS parameter~$\wre$.

Barring the generation of magnetic fields during inflation, the scenario we have
considered in this work involves only standard physics.
Our analysis suggests that the PTA data and consistency with the $\dneff$ bound 
point to a spectrum of PMFs with a small amplitude, but a large spectral index.
The data and constraints also suggest a lower value of~$\wre$.
In fact, the small values of the magnetic field suggested by the PTA data are 
inconsistent with the lower bounds from the Fermi/LAT and HESS observations of 
TeV blazars~\cite{Neronov:2010gir,Tavecchio:2010mk,Dolag:2010ni,Dermer:2010mm,
Vovk:2011aa,Taylor:2011bn,Takahashi:2011ac}.
If we require magnetic fields corresponding to $B_0\simeq 10^{-16}\,\mathrm{G}$
today that are nearly scale-invariant over Mpc scales (as suggested by the CMB
data~\cite{Zucca:2016iur}) and also lead to the SED of GWs indicated by the PTA 
observations, then we may need to consider spectra of PMFs that contain a spectral 
break.
It seems worthwhile to construct inflationary scenarios that naturally lead to 
spectra of PMFs with such a broken power law. 
Since the PMFs affect the tensor perturbations during inflation and reheating over 
large scales, future observations of the B-mode polarization of the CMB can also, 
in principle, help us constrain the parameters describing reheating and reveal 
signatures of inflationary magnetogenesis.
We are presently examining such issues.


\noindent
\underline{\it Acknowledgments:}\/~SM wishes to thank the Council of Scientific 
and Industrial Research, Ministry of Science and Technology, Government 
of India (GoI), for financial assistance.
DM and LS gratefully acknowledge the support received from the Science and
Engineering Research Board, Department of Science and Technology, GoI, through 
the Core Research Grant~CRG/2020/003664.
LS also wishes to thank the Indo-French Centre for the Promotion of Advanced
Research for support of the proposal 6704-4 under the Collaborative Scientific 
Research Programme.
We wish to thank the Gravitation and High Energy Physics Groups at IIT Guwahati 
for illuminating discussions.

\bibliographystyle{apsrev4-2}
\bibliography{references}

\clearpage
\begin{widetext}
\section{Supplementary material}

In this supplementary material, we shall provide a few additional details and 
calculations to support the discussions in the main text.


\subsection{Evolution of the EMFs during reheating}\label{evol_EM}

It is well known that electromagnetic fields are generated during 
inflation by breaking the conformal invariance of the electromagnetic
action (see, for instance, Refs.~\cite{Subramanian:2015lua, Kobayashi:2014sga, 
Haque:2020bip, Tripathy:2021sfb}).
As we had mentioned, we shall assume that the conformal symmetry is 
restored at the end of inflation. 
If we assume that the electrical conductivity of the universe during reheating 
is negligible, both the electric and magnetic fields persist through this period. 
Then, it can be shown that, post inflation, the comoving power spectra of the 
electric and magnetic fields, viz. $\pce$ and $\pcb$, satisfy the equations
\begin{subequations}
\begin{align}
{\pce}'+ {\pcb}' & =0,\\
{\pcb}''+2k^2 (\pcb-\pce) &=0.
\end{align}
\end{subequations}
These coupled equations admit the following exact analytical solutions:
\begin{subequations}\label{eq:pbe}
\begin{align}
\pcb(k,\eta) &=\f{\ae^4}{2} \biggl\{\pbi(k)+\pei(k)
+[\pbi(k)-\pei(k)]\cos[2k(\eta-\ee)]\biggr\},\\
\pce(k,\eta) &= \f{\ae^4}{2} \biggl\{\pbi(k)+\pei(k)
-[\pbi(k)-\pei(k)]\cos[2k(\eta-\ee)]\biggr\},
\end{align}
\end{subequations}
where $\pei(k)$ and $\pbi(k)$ are the spectra of the electric and magnetic fields
at the end of inflation.
When $\pbi(k)>\pei(k)$, on super-Hubble scales such that $k\eta\ll 1$, the above 
solutions simplify to
\begin{subequations}
\begin{align}
\pcb(k,\eta) &\simeq \ae^4\pbi(k),\\
\pce(k,\eta) &\simeq \ae^4 [\pei(k)+\pbi(k)k^2(\eta-\ee)^2].
\end{align}
\end{subequations}
Therefore, during reheating, the power spectra of the electromagnetic fields 
evolve as 
\begin{subequations}
\begin{align}
\pb(k,\eta)&\simeq \l(\f{\ae}{a}\r)^4\pbi(k),\\
\pe(k,\eta) &\simeq \l(\f{\ae}{a}\r)^4
\l[\pei(k)+\pbi(k)\gamma^2(k)\l(\frac{\ae\HI}{aH}-1\r)^2\r],
\end{align}
\end{subequations}
where $\gamma(k)=2(k/\ke)/(1+3\wre)$ and, to arrive at the expression 
for $\pe(k,\eta)$, we have used the relation~\cite{Haque:2020bip}
\begin{align}
k^2(\eta-\ee)^2= \gamma^2(k)
\l(\f{\ae\HI}{aH}-1\r)^2.
\end{align}


\subsection{Upper limit on the magnetic field from the CMB} 

Recall that the energy density of radiation is given by $\rho_\mathrm{r}
=g_\ast \pi^2T^4/30$.
Hence, the dimensionless ratio of the fluctuation in the energy density can 
be expressed as
\begin{equation}
\frac{\delta\rho_\mathrm{r}}{\rho_\mathrm{r}} = 4\f{\delta T}{T}.
\end{equation}
This quantity remains conserved throughout the epoch of radiation domination 
since both the fluctuations and the background evolve as~$a^{-4}$. 
However, it is not conserved during the epoch of reheating. 
During this period, the background energy density evolves as $a^{-3(1+\wre)}$.
Moreover, the fluctuations in the energy density, mainly arising from the energy 
density of the EMFs, evolve as~$a^{-4}$. 
If we assume that all the fluctuations arose from the fluctuations in
the energy density of the EMFs, we can express 
$\delta\rho/\rho$ at the end of the phase of reheating as follows:
\begin{equation}
\l.\f{\delta\rho}{\rho}\r|_{\are} 
= \f{\rhoemi}{\rhoe}\l(\f{\ae}{\are}\r)^{1-3\wre},
\end{equation}
where $\rhoemi$ represents the energy density of the EMFs at the end 
of inflation.

For simplicity, if we neglect the effects after the phase of reheating, we 
can connect the fluctuations in the temperature of the CMB today to the above
quantity as
\begin{equation}
\f{\rhoemi}{\rhoe} 
\l(\f{\ae}{\are}\r)^{1-3\wre} = 4 \l.\f{\delta T}{T}\r|_{\mathrm{CMB}}.
\end{equation}
If we assume that the power spectrum of the magnetic field is dominant 
during inflation, then the quantity $\rhoemi$ is given by
\begin{equation}\label{eq:rhoemi}
\rhoemi = \int_{\kmin}^{\ke}\f{\d k}{k}\pbi(k) 
= \f{\mB^2}{\fnb}, 
\end{equation}
where $\fnb=\nb/[1-(\kmin/\ke)^{\nb}]$ and $\kmin$ is the wave number 
corresponding to the largest scale observable today.

Observations of the CMB constrain the fluctuations in the temperature to be
$\delta T/T \sim 10^{-5}$. 
To avoid a significant impact on the fluctuations in the temperature of
the CMB, the energy density of the PMFs must be small.
Upon using this condition, we obtain that 
\begin{equation}\label{eq:con-mB}
\mathcal{B}^2 = \frac{12 \nb \HI^2 \Mpl^2}{1-(\kmin/\ke)^{\nb}}
\l(\frac{\are}{\ae}\r)^{1-3\wre} \l.\f{\delta T}{T}\r|_{\text{CMB}},
\end{equation}
which leads to 
\begin{align}
\mathcal{B}^2 & \simeq 
12\times 10^{-15}\Mpl^4 \l(\frac{\HI}{10^{-5}\Mpl}\r)^2 \fnb
\quad\times \l(\frac{\are}{\ae}\r)^{1-3\wre}
\times\l(\f{\delta T/T}{10^{-5}}\r)_{\mathrm{CMB}}.
\end{align}

From the definition~(\ref{eq:def_em_pow}) of the power spectrum of the magnetic
field, at the CMB scale of~$1\, \mathrm{Mpc}$, the strength of the magnetic field
{\it today}\/ can be written as
\begin{equation}\label{eq:defb0}
B_0= \mathcal{B}\l(\f{\ae}{a_0}\r)^2
\l(\f{1~\text{Mpc}^{-1}}{\ke}\r)^{\nb/2}.
\end{equation}
Further, upon using Eq.~\eqref{eq:con-mB} in Eq.~\eqref{eq:defb0}, we arrive at
\begin{align}
B_0^{\mbox{\tiny{{CMB}}}}
\simeq 1.17\times 10^{-8}\l(\f{\HI}{10^{-5}\Mpl}\r)\fnb^{1/2}
\times\,\l(\f{1\,\mathrm{Mpc}^{-1}}{\ke}\r)^{\nb/2}
\mathrm{G}.
\end{align}
For a scale-invariant spectrum of the PMFs, the bound 
from fluctuations in the temperature of the CMB turn out to be approximately 
$B_0^{\mbox{\tiny{{CMB}}}} \simeq 1.55$ nG, which is the result we have 
quoted in the text.


\subsection{Spectral density of secondary GWs generated during inflation}
Following the widely accepted models of 
magnetogenesis~\cite{Subramanian:2015lua,Kobayashi:2014sga, Haque:2020bip,
Tripathy:2021sfb,Sorbo:2011rz,Caprini:2014mja,Ito:2016fqp}, we shall assume 
that, towards the end of inflation, the spectrum of the magnetic field is 
of the form $\pb(k,\eta)=\mB^2(-k\eta)^{\nb}$. 
Moreover, we shall assume that the strengths of the electric fields are
considerably weaker.
For simplicity, if we further assume that the background during inflation 
is of the de Sitter form, then the Green's function associated with the 
tensor perturbations at late times is given by $\Gkinf(\eta,\eta_1) \simeq
\eta_1/3$~\cite{Caprini:2014mja}.
Upon substituting these forms for the Green's function and the behavior of 
the electromagnetic modes into Eq.~(\ref{eq:ps-sgws}), we obtain the tensor
power spectrum at the end of inflation to be~\cite{Caprini:2014mja}
\begin{align}
\Psecinf(k,\ee)
&=\f{2\mB^4}{\Mpl^4}
\int_{\kmin}^{\ke}\f{\d q}{q}
\int_{-1}^1 \d\mu \f{(q/k)^{\nb}f(\mu,\beta)}{\l[ 1+(q/k)^2-2\mu(q/k)\r]^{(3-\nb)/2}}
\l[\int_{\eta_i}^{\ee}\d\eta_1 a^2(\eta_1)
\Gkinf(\ee,\eta_1)(k\eta_1)^{\nb} \r]^2\nn\\
&=\f{2\mB^4}{\Mpl^4}\int_{\kmin}^{\ke}\f{\d q}{q}\int_{-1}^1 \d\mu 
\f{(q/k)^{\nb}f(\mu,\beta)}{\l[1+(q/k)^2-2\mu(q/k)\r]^{(3-\nb)/2}}
\l[\int_{x_i}^{\xe}
\f{\d x_1}{3}x_1^{\nb-1}  \r]^2\nn\\
&=\f{2\HI^4}{\Mpl^4}\l(\f{\mB}{\HI^2}\r)^4\mathcal{I}_{\rm inf}^2(k)
\Fnb(k),\label{eq:ptinfs}
\end{align}
where we have set $x_1=k\eta_1$ and the quantities $\Fnb(k)$ and 
$\mathcal{I}_{\mathrm{inf}}(k)$ are given by
\begin{align}
\Fnb(k) 
&=\int_{\kmin}^{\ke}\frac{\d q}{q}\int_{-1}^{1}\d\mu 
\f{(q/k)^{\nb}f(\mu,\beta)}{\l[ 1+(q/k)^2-2\mu(q/k)\r]^{(3-\nb)/2}}
\simeq \f{16}{3\nb}\l[1-\l(\f{\kmin}{k}\r)^{\nb}\r]
+\f{2\alpha}{3}\l[\l(\f{\ke}{k}\r)^{2\nb-3}-1\r],\label{eq:fnb-def}\\
\mathcal{I}_{\mathrm{inf}}(k)
&=\f{1}{3\nb}\l[ \l(\frac{k}{\ke}\r)^{\nb}-\l(\f{k}{k_{*}}\r)^{\nb}\r],
\end{align}
with $\alpha=56/[5(2\nb-3)]$.

In contrast, the tensor power spectrum of the primary GWs originating from 
the vacuum fluctuations during inflation are given by
\begin{equation}\label{pvi}
\Ppriinf(k,\ee)
=\f{2}{\pi^2}\l(\frac{\HI}{\Mpl}\r)^2\,\l(1+\f{k^2}{\ke^2}\r).
\end{equation}


\subsection{Post-inflationary contributions to the secondary tensor power spectrum}

The PMFs are expected to have been generated during inflation due to a 
non-conformal coupling.
We shall assume that the conformal symmetry of the electromagnetic action is 
restored at the end of inflation.
As a result, the electromagnetic modes simply oscillate post-inflation, apart
from the diluting factor that arises due to the expansion of the universe.
The tensor modes $h_{\vk}^{\lambda}(\eta)$ are governed by the equation (see, 
for instance, Refs.~\cite{Sorbo:2011rz,Caprini:2014mja,Sharma:2019jtb})
\begin{align}
h_{\bm k}^{\lambda{}''}+2\f{a'}{a} h_{\bm k}^{\lambda}{}'
+k^2h_{\bm k}^{\lambda}=\mathcal{S}_{\bm k}^\lambda,
\label{eq:tensor-eq-re}
\end{align}
where the source term $\mathcal{S}_{\bm k}^{\lambda}$ is given by
\begin{equation}
\mathcal{S}_{\bm k}^{\lambda}(\eta)
=-\f{2}{\Mpl^2}e^{ij}_{\lambda}({\bm k})
\int \f{\d^3\vq}{(2\pi)^{3/2}} [E_i(\vq,\eta)E_j(\vk-\vq,\eta)
+B_i(\vq,\eta)B_j(\vk-\vq,\eta)].    
\end{equation}
The Green's function corresponding to Eq.~\eqref{eq:tensor-eq-re} satisfies the 
differential equation
\begin{align}
\Gk''(\eta,\eta_1)+2\cH \Gk'(\eta,\eta_1)+k^2 \Gk(\eta,\eta_1)=\delta^{(1)}(\eta-\eta_1).
\end{align}
The Green's function during the epochs of reheating and radiation domination 
can be easily obtained to be
\begin{subequations}
\begin{align}
\Gkre(\eta,\eta_1) 
&= \theta(\eta-\eta_1)\frac{\pi \eta^{l} \eta_1^{1-l}}{2\mathrm{sin}(l\pi)}
\l[J_l(k\eta)J_{-l}(k\eta_1)-J_{-l}(k\eta)J_l(k\eta_1)\r],\label{eq:gre}\\
\Gkra(\eta,\eta_1 )&= \theta(\eta-\eta_1)
\frac{\eta_1}{k\eta}\sin[k(\eta_1-\eta)],\label{eq:gra}
\end{align}
\end{subequations}
where $l=(1-\delta)/2$ with $\delta=4/(1+3\wre)$.

Evidently, at the time of decoupling of the neutrinos~$\eta_\nu$, the tensor 
perturbations generated by the magnetic fields can be expressed as
\begin{align}
h_{\lambda}(\vk,\eta_\nu) =\int_{\ee}^{\ere}\d\eta_1\Gkre(\ere,\eta_1) \Sk(\eta_1)
+\int_{\ere}^{\eta_{\nu}}\d\eta_1\Gkra(\eta_{\nu},\eta_1)\Sk(\eta_1).
\end{align}
Since we have assumed that the electric fields are sub-dominant post-inflation,
we find that the tensor power spectrum [as defined in Eq.~\eqref{eq:ps-sgws}] 
can be expressed as
\begin{align}
\Psec(k,\eta_\nu)
&=\f{2\ae^8}{\Mpl^4}
\l[\int_{\ee}^{\ere}\f{\d\eta_1}{a^2(\eta_1)}\Gkre(\ere,\eta_1)
+\int_{\ere}^{\enu}\f{d\eta_1}{a^2(\eta_1)} \Gkra(\eta_\nu,\eta_1)\r]^2
\int_0^{\infty} \f{\d q}{q}
\int_{-1}^1 \f{\d\mu f(\mu,\beta)\pbi(q)\pbi(|\vk-\vq|)}{[1+\l(q/k\r)^2
-2\mu(q/k)]^{3/2}}\nn\\
&=\f{2\ae^8}{\Mpl^4 k^4}\l[\mCre(\xre,\xe)+\mCra(\xnu,\xre)\r]^2
\int_0^{\infty} \f{\d q}{q}\int_{-1}^1
\f{\d\mu f(\mu,\beta)\pbi(q)\pbi(|\vk-\vq|)}{[1+\l(q/k\r)^2-2\mu(q/k)]^{3/2}}\nn\\
&\simeq\f{2\ae^8}{\Mpl^4 k^4}\l[\mCre^2(\xre,\xe)
+\mCra^2(\xnu,\xre)\r]
\int_0^{\infty} \f{\d q}{q}\int_{-1}^1
\f{\d\mu f(\mu,\beta)\pbi(q)\pbi(|\vk-\vq|)}{[1+\l(q/k\r)^2-2\mu(q/k)]^{3/2}},\label{eq:psec-f}
\end{align}
where we have set $x=k\eta$, and the quantities $\mCre(\xe,\xre)$ and $\mCre(\xe,\xnu)$ are given by
\begin{subequations}
\begin{align}
\mCre(\xre,\xe)
&=k\,\int_{\xe}^{\xre}\f{\d x_1}{a^2(x_1)}\Gkre(\xre,x_1)
=\f{k\xe^{\delta}}{\ae^2}\int_{\xe}^{\xre}\d x_1 x_1^{-\delta}\Gkre(\xre,x_1)
=\f{\xe^{\delta}}{\ae^2} \Cmre(\xre,\xe),\label{eq:mcre}\\
\mCra(\xnu,\xre)
&=k\,\int_{\xre}^{\xnu} \f{\d x_1}{a^2(x_1)}\Gkra(\xnu,x_1)
=\f{k\xre^2}{\are^2}\int_{\xre}^{\xnu}\d x_1x_1^{-2}\Gkra(\xnu,x_1)
=\f{\xre^2}{\are^2}\Cmra(\xnu,\xre).\label{eq:mcra}
\end{align}
\end{subequations}
\color{black}
In arriving at the final expressions above, we have made use of the following forms 
of the scale factor during the epochs of reheating and radiation domination:
\begin{align}
a(\eta)=\l\{\begin{array}{ll}
\ae(x/\xe)^{\delta/2} & \mathrm{for}~\ee \leq \eta \leq\ere,\\
\are (x/\xre) & \mathrm{for}~\ere \leq \eta \leq \eta_{\mathrm{eq}},
\end{array}\r.
\end{align}
where $\eta_\mathrm{eq}$ represents the conformal time corresponding to
the epoch of radiation-matter equality.


\subsubsection{Calculating $\mCre(\xre,\xe)$ and $\mCra(\xnu,\xre)$} 

Let us now compute the quantities $\mCre(\xe,\xre)$ and $\mCre(\xre,\xnu)$
during the epochs of reheating and radiation domination.
During reheating, upon using the functional form~\eqref{eq:gre} for the Green's
function, the indefinite integral in Eq.~\eqref{eq:mcre} can be evaluated to 
be
\begin{align}\label{D2}
\Cmre(x,x_1)
&=  k\int \d x_1 x_1^{-\delta}\Gkre(x,x_1)
=\f{\pi x^l}{2\mathrm{sin}(l\pi)}
\int \d x_1x_1^{1-l-\delta}\l[J_l(x)J_{-l}(x_1)-J_{-l}(x)J_l(x_1)\r]\nn\\
&=2^{-2-l} x^l x_1^{2-2l-\delta}
\biggl\{\f{\Gamma(-l)\Gamma[1-(\delta/2)]}{\Gamma[2-(\delta/2)]}
x_1^{2l}J_{-l}(x)\,
{}_1F_2[1-(\delta/2); 1+l,2-(\delta/2);-(x_1^2/4)]\nn\\
& +4^l \f{\Gamma(l)\Gamma[1-l-(\delta/2)]}{\Gamma[2-l-(\delta/2)]}J_l(x)
{}_1F_2[1-l-(\delta/2);1-l,2-l-(\delta/2);-(x_1^2/4)]\biggr\},
\end{align}
where ${}_1F_2(a,b,c,z)$ denotes the hypergeometric function.
Similarly, upon substituting the Green's function~\eqref{eq:gra} during the 
radiation dominated epoch in Eq.~\eqref{eq:mcra} and carrying out the 
integral, we obtain that 
\begin{align}
\Cmra(\xnu,\xre) 
&=k\int_{\xre}^{\xnu} \d x_1 x_1^{-2}\Gkra(\xnu,x_1)
=\f{1}{\xnu}\int_{\xre}^{\xnu}\f{\d x_1}{x_1} \sin(\xnu-x_1)
=\f{1}{\xnu}\mathcal{I}(\xnu,\xre).
\end{align} 
For convenience, we break the above integral into two domains, i.e. $k<\knu$ 
and $\knu<k<\kre$. 
In these cases, the integral $\mathcal{I}(\xnu,\xre)$ reduces to
\begin{align}
\mathcal{I}(\xnu,\xre)\simeq\l\{\begin{array}{ll}
\l[\gamma_\mathrm{E} 
+\ln\l(k/\kre\r)\r] (k/\knu) & \mbox{for~} k<\knu,\\
\gamma_\mathrm{E} +\ln\l(k/\kre\r) & \mbox{for~} \knu<k<\kre,
\end{array}\r.
\end{align}
where $\gamma_\mathrm{E}=0.577$ is the Euler's constant.


\subsubsection{Contributions to the secondary tensor power spectrum}

At the end of reheating, the secondary tensor power spectrum induced by the magnetic 
fields can be expressed as
\begin{align}
\Psecre(k,\ere) 
&=\frac{2\ae^8}{\Mpl^4k^4} \l[\Cmre(\xe,\xre)\r]^2
\int_0^{\infty} \f{\d q}{q}\int_{-1}^1
\f{\d\mu f(\mu,\beta)\pbi(q)\pbi(|\vk-\vq|}{[1+\l(q/k\r)^2-2\mu(q/k)]^{3/2}}\nn\\
&=\frac{2\HI^4}{\Mpl^4}\l(\frac{\ae^2\mB}{\ke^2}\r)^4
\l(\f{k}{\ke}\r)^{2(\delta+\nb-2)} \l[\Cmre(\xe,\xre)\r]^2\Fnb(k),\label{eq:psec-re}
\end{align}
where $\Fnb(k)$ is given by Eq.~\eqref{eq:fnb-def}.
In a similar fashion, the contribution to the secondary tensor power spectrum due 
to the magnetic fields during the epoch of radiation domination can be expressed as
\begin{align}
\Psecra(k,\enu) 
&=\frac{2\ae^8}{\Mpl^4k^4}\frac{\xre^4}{\are^4}\l[\Cmra(\xre,\xnu)\r]^2
\int_0^{\infty} \f{\d q}{q}\int_{-1}^1
\f{\d\mu f(\mu,\beta)\pbi(q)\pbi(|\vk-\vq|}{[1+\l(q/k\r)^2-2\mu(q/k)]^{3/2}}\nn\\
&=\frac{2\HI^4}{\Mpl^4}\l(\f{\xre}{\xe}\r)^4\l(\frac{\ae}{\are}\r)^4
\l(\frac{\ae^2\mB}{\ke^2}\r)^4\l(\f{k}{\ke}\r)^{2\nb} 
\l[\Cmra(\xre,\xnu)\r]^2\Fnb(k).\label{eq:psec-ra}
\end{align}
Lastly, the contribution due to the cross-term in the secondary tensor power spectrum 
can be expressed as
\begin{align}
\Ptcs(k) 
&=\f{4\ae^8}{\Mpl^4 k^4}\mCre(\xe,\xre)\, \mCra(\xre,\xnu)
\int_0^{\infty} \f{\d q}{q}\int_{-1}^1
\f{\d\mu f(\mu,\beta)\pbi(q)\pbi(|\vk-\vq|}{[1+\l(q/k\r)^2-2\mu(q/k)]^{3/2}}\nn\\
&=\frac{4\HI^4}{\Mpl^4}\l(\f{\ae^2\mB}{\ke^2}\r)^4 \l(\frac{\ae}{\are}\r)^2
\l(\frac{k}{\ke}\r)^{2\nb+\delta-2}\Cmre(\xe,\xre)\,\Cmra(\xre,\xnu)\Fnb(k).
\end{align}
We find that this cross-term is subdominant when compared to the other two terms 
[i.e. those given by Eqs.~\eqref{eq:psec-re} and~\eqref{eq:psec-ra}].
It is for this reason that, in Eq.~\eqref{eq:psec-f}, we have dropped this term 
in the final expression for the secondary tensor power spectrum.


\subsubsection{Spectral shape of $\Omega_{_{\mathrm{GW}}}$
for $k<\kre$ and $k\gg\kre$}

Over wave numbers such that $k\ll\kre$, we can consider the limit wherein 
$\xe\ll \xre \ll 1$.
In such a case, we find that the quantity $\Cmre(\xre,\xe)$ simplifies to be
\begin{align}\label{eq:cm1}
\lim_{k\ll\kre} \Cmre(\xre,\xe)
\simeq\f{1}{(1-\delta)^2} 
\l\{\f{2}{1+2\delta}
-\f{2}{2-\delta} 
\l[1-\l(\f{\kre}{\ke}\r)^{2-\delta}\r]\r\}
\l(\f{k}{\kre}\r)^{2-\delta}.
\end{align} 
Similarly, when $k\gg\kre$, for $\wre>1/3$, the quantity $\Cmre(\xre,\xe)$ 
reduces to
\begin{align}\label{eq:cm2}
\lim_{k\gg\kre}  
\Cmre(\xre,\xe)
&\simeq \sqrt{\f{2}{\pi}}\f{2^{(1-\delta)/2}}{(1-\delta)}
\Gamma[(3-\delta)/2] \Gamma[(\delta-1)/2]\xre^{-\delta/2}
\biggl\{\f{\Gamma[(1-\delta)/2]}{\Gamma(\delta/2)}
\cos\l[\xre-(2-\delta)\pi/4\r]\nn\\ 
& - \frac{2}{2-\delta} 
\cos\l[\xre-\delta\pi/4\r]\biggr\},
\end{align}
whereas, for $\wre<1/3$, the quantity simplifies to
\begin{align}\label{eq:cm3}
\lim_{k\gg\kre} \Cmre(\xe,\xre)
&\simeq  \sqrt{\f{2}{\pi}}\f{2^{(\delta-3)/2}}{(1-\delta)}
\f{\Gamma[(\delta-1)/2]\Gamma[1-(\delta/2)]}{\Gamma[2-(\delta/2)]}
\xre^{-\delta/2}\xe^{2-\delta}  \cos\l[-\xre+\delta\pi/4\r]\nn\\
&\simeq \sqrt{\frac{2}{\pi}}
\frac{2^{(\delta-3)/2}\Gamma[\frac{\delta-1}{2}]}{(1-\delta)(1-\delta/2)}
\xre^{-\delta/2}\xe^{2-\delta}\cos[\xre-(2-\delta)\pi/4].
\end{align}

At the end of reheating, the power spectrum of secondary GWs generated
due to the magnetic fields [defined in Eq.~\eqref{Psec}] is given by
\begin{align}\label{eq:def_psec}
\Psecre(k,\ere)    
= \f{2\HI^4}{\Mpl^4} \l(\f{\ae^2\mB}{\ke^2}\r)^4 
\l(\frac{k}{\ke}\r)^{2(\delta+\nb-2)}\,\Fnb(k)\, \mathcal{C}^2(\xre,\xe).
\end{align}
On substituting Eq.~\eqref{eq:cm1} in this expression, for $k\ll \kre$,
we obtain that
\begin{align}
\Psecre(k,\ere)  
\simeq \mathcal{A}_1 \l(\f{\mB \ae^2}{\ke^2}\r)^4 \l(\f{\HI}{\Mpl}\r)^4
\l(\f{\kre}{\ke}\r)^{2(\delta-2)} \l(\frac{k}{\ke}\r)^{2\nb} \Fnb(k).
\end{align}
Similarly, on utilizing Eqs.~\eqref{eq:cm2} and \eqref{eq:cm3}, 
for $k > \kre$, we obtain the following expressions
\begin{subequations}
\begin{align}
\Psecre(k,\ere) 
&\simeq \mathcal{A}_2 \l(\f{\mB \ae^2}{\ke^2}\r)^4
\l(\f{\HI}{\Mpl}\r)^4 \l(\f{\kre}{\ke}\r)^{2(\delta-2)}
\l(\frac{k}{\kre}\r)^{-2-|\nw|}\l(\frac{k}{\ke}\r)^{2\nb} \Fnb(k),\\
\Psecre(k,\ere) 
&\simeq \mathcal{A}_3 \l(\frac{\mB \ae^2}{\ke^2}\r)^4
\l(\f{\HI}{\MP}\r)^4
\l(\f{\kre}{\ke}\r)^{2(\delta-2)}
\l(\f{k}{\kre}\r)^{-2-|\nw|}\l(\frac{k}{\ke}\r)^{2\nb} \Fnb(k),
\end{align}
\end{subequations}
when $\wre>1/3$ and  $\wre<1/3$, respectively.
The quantities $\mathcal{A}_1$, $\mathcal{A}_2$ and $\mathcal{A}_3$ 
that appear in the above expressions are given by
\begin{subequations}
\begin{align}
\mathcal{A}_1 
&=\f{2}{(1-\delta)^4} 
\l\{\f{2}{1+2\delta}
-\f{2}{2-\delta} 
\l[1-\l(\f{\kre}{\ke}\r)^{2-\delta}\r]\r\}^2,\\
\mathcal{A}_2 
&=\f{2^{(3-\delta)}\Gamma^2[(3-\delta)/2]
\Gamma^2[(\delta-1)/2]}{\pi(1-\delta)^2}
\biggl\{\f{\Gamma[(1-\delta)/2]}{\Gamma(\delta/2)}
\cos\l[(k/\kre)-(2-\delta)\pi/4\r] 
-\f{2}{2-\delta} 
\cos\l[(k/\kre)-\delta\pi/4\r]\biggr\}^2,\\
\mathcal{A}_3 
&=\f{2^{\delta-1}\Gamma^2[(\delta-1)/2]}{\pi(1-\delta)^2(1-\delta/2)^2}
\l(\f{\kre}{\ke}\r)^{2(2-\delta)}\cos^2[(k/\kre)-\delta{\pi}/{4}].
\end{align}
\end{subequations}


\subsection{Comparison with the PTA data}

In this section, we shall briefly discuss the constraints arrived at by 
comparing our scenarios with the PTA data. 
We explore different combinations of the parameters involved, viz. $(\mB,
\nb,\wre,\Tre)$, to identify the best-fit values. 
We have carried out six independent runs for the range of parameters 
listed in Tab.~\ref{tab:pe}.
\begin{table}[!h]
\centering
\begin{tabular}{|c|c|c|c|c|c|c|}
\hline
Model & Parameter & Prior & Mean value
& $\mathcal{B}_{\rm X, Y}$ & $B_0\,(1\,\text{Mpc}^{-1})\, \mathrm{G}$ & $\dneff$ Bound\\
\hline
R1 & $\wre$ & (0,1.0)& $0.16^{+0.18}_{-0.10}$ & & &\\
& $\log_{10}(\Tre)$ & (-2,0) & $-0.55^{+0.27}_{-0.09}$ &  & &\\
& $\nb$ & (0,1.5) & $1.1^{+0.15}_{-0.27}$ & $33.38\pm 7.8$ & $4.12\times 10^{-22}$& $\times$\\
& $\Tilde{\mathcal{B}}$ & (-3,3) & $-0.95^{+0.73}_{-0.82}$ & & & \\
\hline
R2 & $\wre$ & 0.333 & 0.333 & & &\\
& $\log_{10}(\Tre)$ & (-2,0) & $-0.55^{+0.27}_{-0.09}$ &  & &\\
& $\nb$ & (0,1.5) & $0.90^{+0.21}_{-0.20}$ & $15.36\pm 4.43$ & $3.34\times 10^{-16}$ & $\times$\\
& $\Tilde{\mathcal{B}}$ & (0,5) & $3.10^{+0.73}_{-0.82}$ & & &\\
\hline
R3 & $\wre$ & 0.0 & 0.0 & & &\\
&  $\log_{10}(\Tre)$ & $(-2,0)$ & $-0.51^{+0.27}_{0.09}$ & & & \\
& $\nb$ & (0,1.5) & $0.91^{+0.21}_{-0.20}$ & $18.65\pm 7.34$ & $2.53\times 10^{-25}$& \checkmark\\
& $\log_{10}(\Tilde{\mathcal{B}})$ & (0,5) & $3.03^{+0.77}_{-0.62}$ &  & &\\
\hline     
R4 & $\wre$ & 0.1 & 0.1 & & &\\
& $\log_{10}(\Tre)$ & (-2,1) & $-0.57^{+0.47}_{-0.17}$ &  & &\\
& $\nb$ & (0,0.5) & $0.^{+0.21}_{-0.20}$ & $15.36\pm 4.43$ & $8.7\times 10^{-18}$& \checkmark\\
& $\Tilde{\mathcal{B}}$ & (0,5) & $1.96^{+0.17}_{-0.82}$ & & &\\
\hline      
R5 & $\wre$ &  $(0,0.333)$ & $0.14^{+0.12}_{-0.09}$ & & &\\
& $\Tre$ & $0.25\,\GeV$ & $0.25\,\GeV$ & & &\\
& $\nb$ & $(0,1.5)$ & $1.1^{+0.21}_{-0.20}$ & $13.35\pm 3.47$ & $3.4\times 10^{-22}$ & $\times$\\
& $\Tilde{\mathcal{B}}$ & $(0,7)$ & $2.74^{+0.76}_{-0.82}$ & & &\\
\hline 
R6 & $\wre$ & $0,1.0$ & $0.16^{+0.11}_{-0.09}$& & &\\
& $\Tre$ & $0.1$ & $0.1$ &  & &\\
& $\nb$ & $(0,1.0)$ & $0.94^{+0.05}_{-0.08}$ & $11.59\pm 3.51$ &$8.8\times 10^{-21}$ & $\times$\\
& $\tilde{\mathcal{B}}$ & $(0,5)$ & $2.89^{+0.38}_{-0.78}$ & & &\\
\hline 
\end{tabular}
\caption{We have listed the priors on the parameters, the best-fit values
and the Bayesian evidence ($\mathcal{B}_{\rm X,Y}$), arrived at from the 
six MCMC runs we have carried out comparing the scenario with the NANOGrav 
15-year data.
In addition, we have listed the strength of the magnetic field today that
corresponds to the best-fit values.}\label{tab:pe}
\end{table}
In the table, we have listed the best-fit values and the Bayesian evidence 
($\mathcal{B}_{\rm X,Y}$) for the scenario when compared to the SMBHB model.
We have also indicated the current strength of the magnetic field in the 
scenario.
In Figs.~\ref{fig:post1} and~\ref{fig:post2}, we have illustrated the 
posterior distributions for the runs R1, R4, R5 and R6.
After arriving at the best-fit values, we have also checked whether the total
SED of primary and secondary GWs are consistent with the $\dneff$ bound.
We find that the PTA data, when combined with the $\dneff$ bound, suggests 
that $\wre < 1/3$. 
In fact, the most preferred EoS turns out to be $\wre = 0$ with a reheating 
temperature of $\Tre \simeq 1~\text{GeV}$. 
This scenario has a strong Bayesian evidence, with $\mathcal{B}_{\rm X,Y} 
= 18.65 \pm 7.34$.  
\begin{figure}[!t]
\includegraphics[width=0.45\linewidth]{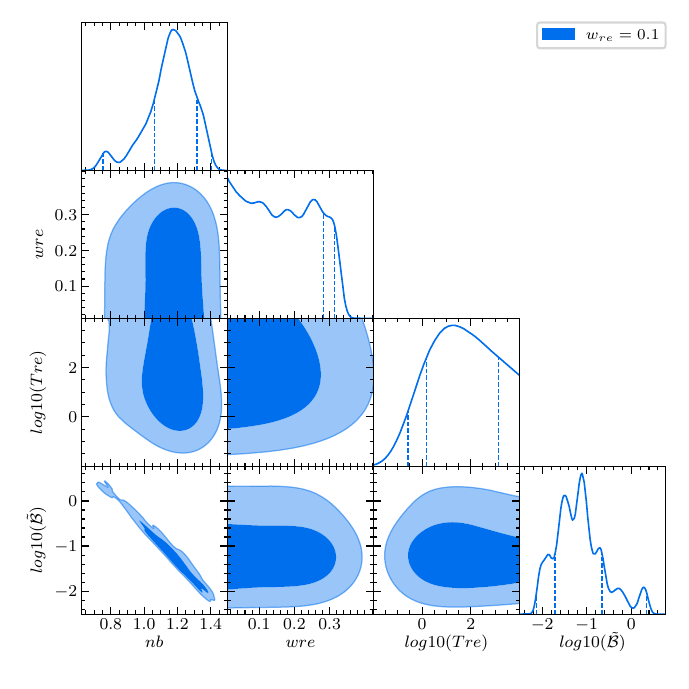}
\includegraphics[width=0.45\linewidth]{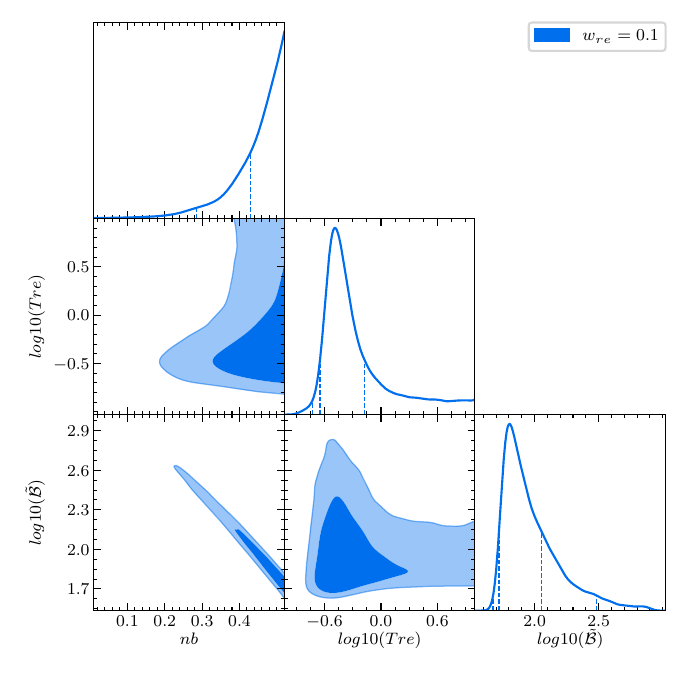}
\caption{We have presented the posterior distributions for the parameters 
in the cases of the runs R1 (on the left), and R4 (on 
the right).}\label{fig:post1}
\end{figure}
\begin{figure}[!t]
\includegraphics[width=0.475\linewidth]{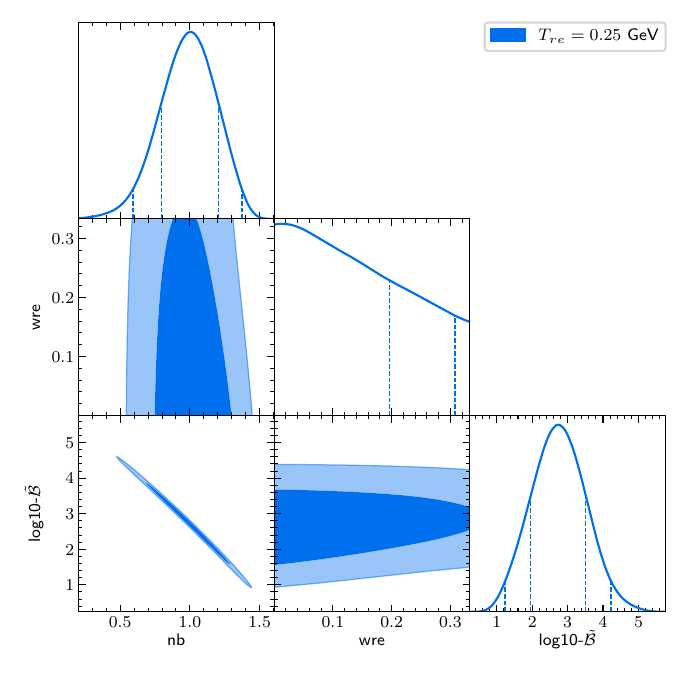}
\includegraphics[width=0.475\linewidth]{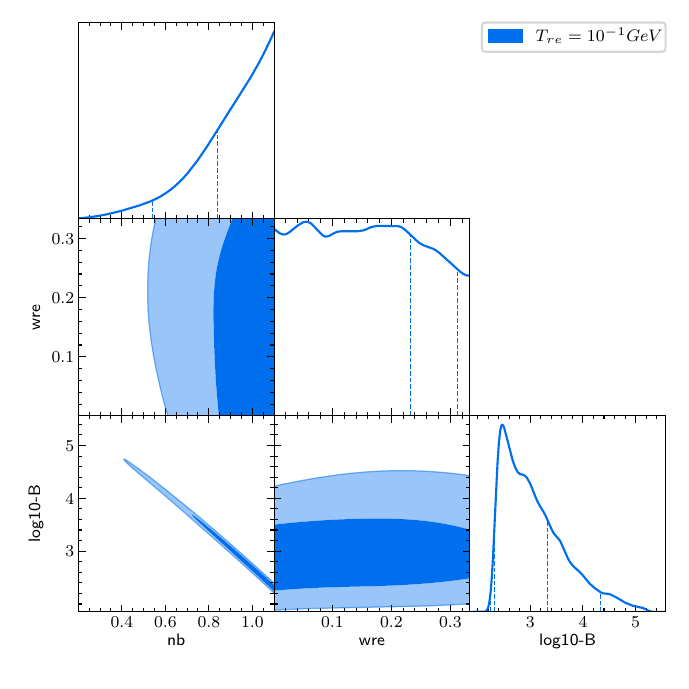}
\caption{The posterior distributions for the parameters in the cases of the 
runs R5 (on the left) and R6 (on the right).}\label{fig:post2}
\end{figure}

\end{widetext}


\end{document}